\documentclass[12pt,a4paper]{article}
\usepackage{fullpage}
\usepackage{amssymb}
\usepackage{amsmath}

\newtheorem{theoreme}{Theorem}[section]
\newtheorem{prop}[theoreme]{Proposition}
\newtheorem{cor}[theoreme]{Corollary}
 
\newtheorem{lemme}[theoreme]{Lemma}

\newcommand{\card}{{\rm card}}
\newcommand{\R}{\mathbb{R}}
\newcommand{\Q}{\mathbb{Q}}

\def\C{\mathbb{C}}
\newcommand{\Z}{\mathbb{Z}}
\newcommand{\N}{\mathbb{N}}

\newcommand{\points}{{\cal M}_{\sigma}}

\def\denssup#1{\overline{\rm dens}\left(#1\right)}
\def\od{\overline{d}}

\renewcommand{\epsilon}{\varepsilon}

\begin{document}

\title{Quasicrystals and almost periodicity}

\author{Jean-Baptiste Gou\'er\'e\footnote{\textit{Postal address}: Universit\'e Claude 
Bernard Lyon 1,
LaPCS, B\^atiment B, Domaine de Gerland, 50, avenue Tony-Garnier F-69366 Lyon 
Cedex 07, France.
\textit{E-mail}: jbgouere@univ-lyon1.fr }}  

\date{}

\maketitle

\begin{abstract}
We introduce  a topology ${\cal T}$ on the space $U$ of uniformly discrete subsets of the Euclidean space.
Assume that $S$ in $U$ admits a unique autocorrelation measure.
The diffraction measure of $S$ is purely atomic if and only if $S$ is almost periodic in $(U,{\cal T})$.
This result relates idealized quasicrystals to almost periodicity.
In the context of ergodic point processes, the autocorrelation measure is known to exist.
Then, the diffraction measure
is purely atomic if and only if the dynamical system has a pure point spectrum.
As an illustration, we study deformed model sets. 

\end{abstract}

\section*{Introduction}
From a physical point of view, the  diffraction of X rays by a material is a way of investigating its microscopic structure.
The interferences of X rays that are diffracted by different atoms are indeed linked to the relative positions of
these atoms. If the atoms form a lattice, as in ideal crystals,
then the diffraction pattern consists in sharp peaks.
Quasicrystals, discovered in $1982$ by Shechtman et al.\ \cite{SBGC}, have the same kind of diffraction
pattern, but they are not periodic. 
That raises the question of understanding which kind of atoms distribution
produces a diffraction pattern consisting in sharp peaks. 

In Section~\ref{s.1}, we recall some results about locally finite sets and uniformly discrete sets, and we  define some pseudo-metrics on the space of uniformly discrete sets. 
Section~\ref{s.2} collects some known results about almost periodicity. 
In Section~\ref{s.3}, we recall the mathematical formalism of diffraction,
and we state our characterization of pure point diffraction in terms of almost periodicity.
As an application, we treat in Section~\ref{s.4}
the example of deformed model sets.

\section{Locally finite sets and uniformly discrete sets}
\label{s.1}
\subsection{Locally finite sets}

%Let $\points$ denote the set of locally finite subsets of $\R^n$ with $d\ge1$. 
Following \cite{Solomyak-spectrum-delone-set}, we define a metrics $d$ on the set $\points$
 of locally finite subsets of $\R^n$, by
$$
d(S,S')=\min\left\{1/\sqrt{2}, \inf D(S,S')\right\},
$$
where
$$
D(S,S')=\{a>0 \; : \; S\cap B_{1/a} \subset S'+B_a \hbox{ and }S'\cap B_{1/a} \subset S+B_a\}.
$$
Note that $D(S,S')$ is a half line, unbounded to the right, i.e., that 
$D(S,S')$ contains $[a,+\infty[$ as soon as $D(S,S')$ contains $a$.

The topology ${\cal T}$ generated by $d$ can also be defined as follows, see \cite{Matheron}.
If $K$ is a compact subset of $\R^n$, define
$${\cal F}^K = \{S \in \points : S\cap K = \emptyset\}.$$
If $O$ is an open subset of $\R^n$, define
$${\cal F}_O = \{S\in \points : S \cap O \neq \emptyset\}.$$
Then, ${\cal T}$ is also the topology generated by the union of the 
two families ${\cal F}^K$, $K$ compact,
and ${\cal F}_O$, $O$ open.

For $t \in \R^n$ let $T_t:\points\to\points$ denote the translation defined by
$$T_t (S)=S-t=\{x-t\,:\,x\in S\}.$$
Translations are homeomorphisms of $\points$.
We now define a $\sigma$-algebra ${\cal A}$ on $\points$.
For $A\subset\R^n$, let $N_A:\points\to\N\cup\{\infty\}$ denote
\begin{equation} \label{ppna}
N_A(S)= \card(A \cap S). 
\end{equation}
The $\sigma$-algebra ${\cal A}$ on $\points$ is generated by the family of maps $N_A$, for every Borel set$A$, or, equivalently, for every compact set $A$.
Note that $\cal A$ is also the Borel $\sigma$-algebra of $(\points,{\cal T})$.

\subsection{Uniformly discrete sets}

A set $S \in \points$ is uniformly discrete with parameter $r>0$ if two different points of $S$ are at least at distance $r$.
The set $U_r$ of such  $S \in \points$ inherits the metric structure of $(\points,d)$.
The space $(U_r,d)$ is compact.

We now define a new topology on $U_r$, by specifying various uniformly equivalent pseudo-metrics that induce this topology.
These pseudo-metrics are connected to Besicovitch pseudo-metrics on spaces of functions defined by formulae of the kind
$$d_B(f,g)=\limsup_{R\to\infty} \frac{1}{|B_R|}\left(\int_{B_R}|f-g|^p)\right)^{1/p}$$
where $B_R$ is the ball of radius $R$ of $\R^n$ centered at the origin and $|B_R|$ its volume.
%Firstly, we give the two following definitions.
%We need two definitions.

If $S \in \points$, its upper density is defined by
$$\denssup{S}=\limsup_{R \rightarrow \infty}  \frac{1}{|B_R|}\card(S \cap B_R).$$
Finally, call a set $A \subset \R^n$ relatively dense if there exists $M>0$ such that each ball of $\R^n$
of radius $M$ meets $A$.

\noindent $\bullet$ The pseudo-metrics $\od$. For $a>0$ and $S$, $S'$ in $U_r$, define
$$
\Delta_a(S,S')=\widetilde{\Delta}_a(S,S') \cup \widetilde{\Delta}_a(S',S),
$$
where
$$\widetilde{\Delta}_a(S,S')=\{x \in S : (x+B_a) \cap S' = \emptyset\}.$$
Let  $\overline{D}(S,S')$ be the set of all real $a>0$ such that 
$$
\denssup{\Delta_a(S,S')}\leq a.
$$
Note that $\overline{D}(S,S')$ is a half line, unbounded to the right. Define
$$\overline{d}(S,S')=\min\left\{r/2,\inf(\overline{D}(S,S'))\right\}.$$

\noindent $\bullet$  The pseudo-metric $\od^c$.
For $S$, $S'$ in $U_r$, define
$$\od^c(S,S') = \limsup_{R\to\infty}\frac{1}{|B_R|} \int_{B_R} d(S-t,S'-t)dt.$$

\noindent $\bullet$ The pseudo-metric $\od_f$. Let $f:\R^n\to\R$ be continuous
with support in $B_{r/5}$.
We assume that $f$ is not everywhere equal to $0$. 
Notice that the function
$$\begin{cases}
(U_r,d) & \rightarrow \R \\
S & \mapsto \mu_S*f(0),
\end{cases}$$
where $\mu_S=\sum_{x \in S} \delta_x$, is continuous, and 
that, for all $u \in \R^n$, $\mu_S*f(u)=\mu_{S-u}*f(0)$.
For $S$ and $S'$ in $U_r$, define
$$
\overline{d}_f(S,S')=\limsup_{R \rightarrow \infty} \frac{1}{|B_R|} \int_{B_R} |\mu_S*f(u)-\mu_{S'}*f(u)|du.
$$

We prove the following proposition:

\begin{prop} The three functions $\od$, $\od^c$ and $\od_f$ define translation invariant pseudo-metrics.
These pseudo-metrics are uniformly equivalent. Therefore, they define the same topology and
the same Cauchy sequences.
\end{prop}

To show the uniform continuity of the identity map from $(U_r,\od^c)$ to $(U_r,\od_f)$, we use and prove the following, stronger, result:

\begin{lemme} \label{mfcontinue} Let $H:(U_r,d)\to\R$ be continuous. 
Then, for all $\epsilon>0$,
there exists $\eta>0$ such that, for every $S$ and $S'$ in $U_r$ such that $\od^c(S,S') \leq \eta$,
$$
\limsup_{R\to\infty} \frac{1}{|B_R|} \int_{B_R} |H(S-t)-H(S'-t)|dt \leq \epsilon.
$$
\end{lemme}
%
%Let $(\udr,\od)$ denote the quotient space obtained by identifying $S$ and $S'$ in  $(U_r,\od)$ if 
%and only if $\od(S,S')=0$.
We get the following result:
\begin{prop} The quotient space of $U_r$ by the pseudometric $\od$ is a complete metric space.
\end{prop}

\section{Almost periodicity}
\label{s.2}

Let $E$ be a Banach space, and $f:\R^n\to E$. For $t \in \R^n$, let $f_t$ be
defined by $f_t(x)=f(x-t)$. 
The function $f$ is said to be Bohr almost periodic if, for all $\epsilon>0$,
the set
$$P_{\epsilon}=\{t \in \R^n:\|f-f_t\|_{\infty}<\epsilon\}$$
is relatively dense (see for example \cite{Levitan}).

The following result is classical and proved in a more general context in 
\cite{Eberlein-fourier-pp,Hewitt-fourier-pp,Segal-fourier-pp}.

\begin{theoreme} \label{mesure-pp-fini} 
Let $\mu$ be a finite measure on $\R^n$. Then, the two following assertions are equivalent:
\begin{enumerate}
\item The function $\widehat{\mu}$ is Bohr almost periodic.
\item The measure $\mu$ is purely atomic.
\end{enumerate}
\end{theoreme}
This can be extended to classes of unbounded measures (see for example \cite{Gil-de-Lamadrid-al-pp}).
The following result, which is an easy consequence of Theorem \ref{mesure-pp-fini}, is sufficient for our purposes:

\begin{theoreme} \label{mesure-pp-general} Let $\mu$ be a positive, tempered and positive definite measure on $\R^n$.
Let $f$ be a Schwartz function from $\R^n$ to $\C$ such that $\widehat{f}(x)>0$ for all $x\in\R^n$.
Then, the two following assertions are equivalent:
\begin{enumerate}
\item $\mu*f$ is Bohr almost periodic.
\item $\widehat{\mu}$ is purely atomic.
\end{enumerate}
\end{theoreme}
We need the following extension of the definition of almost periodicity.
Let $(X,d_0)$ be a pseudometric space. We consider an action of $\R^n$ on $X$ such that
\begin{enumerate}
\item For all $t \in \R^n$, $x \mapsto x-t$ is an isometry.
\item For all $x \in E$, $t \mapsto x-t$ is continuous.
\end{enumerate}
Say that $x\in X$ is almost periodic if, for all $\epsilon>0$, the set
$$\{t \in \R^n : d_0(x-t,x) \leq \epsilon\}$$ 
is relatively dense. For $x\in E$, let $O_x=\{x-t, t\in\R^n\}$.
Recall that $A\subset E$ is totally bounded if, for all $\epsilon>0$, $A$ 
can be covered by a finite number of balls of radius $\epsilon$
centered at points of $A$.
Recall that:

\begin{theoreme} \label{pp-relcomp} The point 
$x$ is almost periodic if and only if $O_x$ is totally bounded. 
\end{theoreme} 

\section{Diffraction}
\label{s.3}
\subsection{Generalities}

We now recall the mathematical formalism of diffraction theory, following Hof~\cite{Hof-95}.
Let $S \in \points$.
For a given $R>0$, we define the measure $\gamma_R$ by
\begin{equation} \label{def-ac-points}
\gamma_R:=\frac{1}{|B_R|} \sum_{x,y \in S \cap B_R} \delta_{y-x}
=\frac{1}{|B_R|} \; \left(\sum_{x \in S \cap B_R} \delta_x\right)*\left(\sum_{x \in S \cap B_R} \delta_{-x}\right).
\end{equation}
%where $\delta_x$ is the Dirac measure at $x$ and $|B_R|$ is the canonical Lebesgue measure of the ball $B_R$. 
Thus, $\gamma_R$
represents the relative positions of the points of $S\cap B_R$.
Physically, the Fourier transform of $\gamma_R$ corresponds with the diffraction pattern: 
$\widehat{\gamma_R}(t)$ is the luminous intensity diffracted in the direction $t$ by a material whose atom centers
are the points of $S\cap B_R$.

If it exists, the limit $\gamma$ in the vague topology as $R \rightarrow \infty$ of the measures $\gamma_R$
is called the unique autocorrelation of $S$.
This measure is tempered and its Fourier transform $\widehat{\gamma}$ is a positive measure, called the
diffraction measure of $S$.

\bigskip

We prove the following lemma:

\begin{lemme} \label{ac-plouf}
Let $\psi:\R^n\to\R$ be continuous and nonnegative with compact support.
Assume that the integral of $\psi$ is equal to $1$.
If $f:\R^n\to\R$ is continuous with compact support, let
$H_f:\points\to\R$ be defined by
$$H_f(S)=\sum_{x,y \in S} \psi(x)f(y-x).$$
Then, the restriction of $H_f$ to $(U_r,d)$ is continuous.
Let $\gamma$ be a positive and locally finite measure on $\R^n$. 
Then,  $\gamma$ is the unique autocorrelation of $S\in \points$ if and only if, for all $f:\R^n\to\R$ continuous with compact support,
$$
\frac{1}{|B_R|}\int_{B_R} H_f(S-t)dt \longrightarrow \gamma(f), \quad R \longrightarrow \infty.
$$
Therefore, $S$ admits a unique autocorrelation if and only if $\displaystyle\frac{1}{|B_R|}\int_{B_R} H_f(S-t)dt$
converges for all $f$ continuous with compact support.
\end{lemme}
Let $A_r$ denote the set of all the elements $S$ of $U_r$ with a unique autocorrelation.
Lemmas \ref{ac-plouf} and \ref{mfcontinue} yield:
\begin{prop} \label{ac-ferme} The set $A_r$ is a closed subset of $(U_r,\od)$.
\end{prop}

\subsection{Patterson sets}

Following \cite{Lagarias-math-qc}, $S\in A_r$ is a Patterson set if its diffraction measure is 
purely atomic (actually,
\cite{Lagarias-math-qc} also requires  $S$ to be relatively dense). A Patterson set is a mathematical
idealization of a quasicrystal.

Using Theorem \ref{mesure-pp-general}, we prove the following result:

\begin{theoreme} \label{caract-qc} Let $S\in A_r$ with unique autocorrelation measure $\gamma$.
The following assertions are equivalent:
\begin{enumerate}
\item $S$ is a Patterson set.
\item $\widehat{\gamma}$ is purely atomic.
\item For all $R>0$ and $\epsilon>0$, the set
$\{t\in\R^n:\gamma(t+B_R)\geq\gamma(0)-\varepsilon\}$
is relatively dense.
\item $S$ is almost periodic in $(U_r,\od)$.
\item For all $\epsilon>0$, the set
$\left\{t \in \R^n : \denssup{\widetilde{\Delta}_{\epsilon}(S,S-t)} \leq \epsilon\right\}$
is relatively dense. 
\end{enumerate}
\end{theoreme}
Theorem \ref{caract-qc} and Proposition \ref{ac-ferme} yield
\begin{cor} The Patterson sets form a closed subset of $(U_r,\od)$.
\end{cor}
Here is a corollary of Theorem \ref{caract-qc}:
\begin{cor}  \label{tropdepomme}  Let $S\in A_r$ with unique autocorrelation measure $\gamma$.
\begin{enumerate}
\item If, for all $\epsilon>0$, the set 
$$\{t \in \R^n : \gamma(t)\geq\gamma(0)-\epsilon\}$$
is relatively dense, then $S$ is a Patterson set. The converse is true if $S-S$ is uniformly discrete.
\item If $S$ is almost periodic in $(U_r,\widetilde{d})$, where $\widetilde{d}$ is the pseudo-metric defined by
$$
\widetilde{d}(S,S')=\denssup{S\Delta S'},
$$
then $S$ is a Patterson set. The converse is true if $S-S$ is uniformly discrete.
\end{enumerate}
\end{cor}
The first point of Corollary \ref{tropdepomme} is proved in \cite{Baake-Moody-poids}.
In that paper, the point sets can be weighted subsets of a  locally compact Abelian group.
The sufficient condition for being a Patterson set given by the second point of Corollary \ref{tropdepomme}
is proved in \cite{Solomyak-spectrum-delone-set} under unique ergodicity assumptions.

\subsection{Point processes}

A point process \cite{Moller,Neveu-pp} is a measurable map $\chi$ from a probability space
$(\Omega,\cal{F},P)$ to $(\points,\cal{A})$.
A point process $\chi$ is uniformly discrete if there exists $r>0$ such that $\chi$ takes its values in $U_r$.
A point process is (square) integrable if, for
every compact set $A$, the random variable $N_A$ defined by (\ref{ppna}) is (square) integrable. 
%It is said square integrable if the variables $N_A$ are square integrable.
A point process is stationary if its law is invariant under the action of the translations
$(T_t)_{t \in \R^n}$.
Let $\chi$ denote a stationary and integrable point process, and $\gamma_R$ 
denote the random measure associated with $\chi$ by (\ref{def-ac-points}).
The Palm measure of $\chi$ is the 
measure $\widetilde{P}$ on $(\points,\cal{A})$ defined by 
$$
\widetilde{P}(F) := \frac{1}{|B|}E\left(\sum_{x \in \chi \cap B} 1_F(\phi-x)\right),
$$
for $F \in \cal{A}$,
where $B$ is a fixed Borel subset of $\R^n$ whose Lebesgue measure $|B|$ is finite and strictly positive, see \cite{Moller}.
This definition does not depend on $B$.

Let ${\cal B}(\R^n)$ denote the Borel $\sigma$-algebra of $\R^n$.
If $m$ is a measure on $(\points,{\cal A})$, its intensity $I(m)$ is the measure on  
$(\R^n,{\cal B}(\R^n))$ 
defined by
$$I(m)(A)=\int \card(\phi \cap A) \; dm(\phi), \quad  A \in {\cal B}(\R^n).
$$
%where ${\cal B}(\R^n)$ is the set of Borel subsets of $\R^n$.

\medskip

The following result is proved in \cite{G-diffraction-palm}:

\begin{theoreme} \label{ac-palm}
Let $\chi$ be a stationary ergodic and square integrable point process on
$\mathbb{R}^d$. Let $J$ be the (deterministic) intensity of the Palm measure of $\chi$.
Then, $J$ is locally finite and
for each bounded Borel set $A$ in $\R^n$,
$$\lim_{R \rightarrow \infty} \gamma_R(A) = J(A)  \;\;\; a.s.$$
In particular, $\chi$ admits a.s.\ $J$ as a unique autocorrelation measure.
\end{theoreme}
Actually, we can prove the following result:
\begin{theoreme} \label{ipalmpatterson} Let $\chi$ be a stationary and uniformly discrete point process on $\R^n$, and $J$ the intensity of its Palm measure.
If $\widehat{J}$ is purely atomic, then $\chi$ is a.s. a Patterson set. The converse is true if $\chi$
is ergodic.
\end{theoreme}
Using Theorem \ref{mesure-pp-general}, we prove the following result:
\begin{theoreme} 
\label{ipalmat} 
Let $\chi$ be a uniformly discrete and stationary point process, and $J$ the intensity of its Palm measure.
The following assertions are equivalent:
\begin{enumerate}
\item 
$\widehat{J}$ is purely atomic.
\item 
For all $R>0$ and for all $\epsilon>0$, the set
$$
\left\{t \in \R^n \; : \; J(\{0\})-J(B_R+t)\leq \epsilon\right\}
$$ 
is relatively dense.
\item 
For all $R_0>0$, there exists $R$ in $]0,R_0[$ such that, for all $\epsilon>0$, the set
$$
\left\{t \in \R^n \; : \; P\Big(\{\chi \cap B_R \neq \emptyset \} \: \Delta \:  \{(\chi-t) \cap B_R \neq \emptyset \}\Big)
\leq \epsilon\right\}
$$ 
is relatively dense.
\item 
There exists ${\cal G}$ a generating subset of  ${\cal A}$ such that, for all $A \in {\cal G}$ and for all $\epsilon>0$,
the set
$$
\left\{t \in \R^n \; : \; P(A \: \Delta \:  (A+t)) \leq \epsilon\right\}
$$
is relatively dense.
\item 
For all $A$ in ${\cal A}$ and for all $\epsilon>0$, the set
$$
\left\{t \in \R^n \; : \; P(A \: \Delta \:  (A+t)) \leq \epsilon\right\}
$$
is relatively dense.
\item 
For every square integrable $f$ on $\points$, $\phi^f:t \mapsto f_t$ is Bohr almost periodic.
\item 
The dynamical system $(\points,P,{(T_t)}_t)$ admits a discrete spectrum. 
\end{enumerate}
\end{theoreme}
Related results are proved in \cite{Lee-Moody-Solomyak-pp} under finite local complexity assumptions. See also \cite{Solomyak-spectrum-delone-set}.
\begin{cor} Let $\chi$ be a uniformly discrete and stationary point process, and $J$ the intensity of its Palm measure.
If, for all $\epsilon>0$, the set
$$\{t\in\R^n:J(\{t\})\geq J(\{0\})-\epsilon\}$$
is relatively dense, then $\widehat{J}$ is purely atomic. The converse is true if
the support of $J$ is uniformly discrete.
\end{cor}

\section{Example: deformed model sets}
\label{s.4}
In this section, we provide an example of application of our results.
Deformed model sets are studied in \cite{Bernuau-Duneau}. 
Let $E$ and $F$ be two linear subspaces of $\R^n$ such that $\R^n=E \oplus F$.
If $x \in \R^n$ we write $x=x_E+x_F$ with obvious notations.
We denote by $B_R^E$ (resp. $B_R^F$) the closed ball of $E$ (resp. $F$) centered at the origin and of radius $R$.
Let $|\cdot|_F$ denote the canonical Lebesgue measure on $F$.
We assume that $E$ contains a vector $v$ whose coordinates are linearly independent over $\Q$.
Let W be a fixed bounded Borel subset of $F$. 
Let $g:W\to E$ be uniformly continuous.
Let ${\mathbb T}^n=\R^n/\Z^n$ denote the $n$ dimensional torus, and $\phi:{\mathbb T}^n\to{\cal P}(E)$ be defined by
$$\phi(u)=\psi((E+W) \cap (u+\Z^n)),
$$
where $\psi:E+W\to E$ is defined by
$$
\psi(z)=z_E+g(z_F).
$$
For all $t\in E$,
$\phi(u-t)=\phi(u)-t.$
We use the vector space $E$ to define $U_r$ as in Section~\ref{s.1}, and we 
assume that there exists an $r>0$ such that
$$\phi({\mathbb T}^n)\subset U_r.$$
Then, $\phi$ is measurable from ${\mathbb T}^n$ endowed with its natural Borel $\sigma$-algebra to $(U_r,{\cal A})$.

\subsection{Deterministic results}

In this subsection, we assume that $|\partial W|_F=0$.
We prove the following lemma:

\begin{lemme} \label{dms-riemann}
The map $\phi:{\mathbb T}^n\to(U_r,d)$ is continuous on the complementary of a set of vanishing
Haar measure.
\end{lemme}

With that lemma the following results are simple consequences of the uniform ergodicity of $U_r$ under
the action of the translations ${(T_t)}_{t\in E}$.

\begin{prop} \label{dms-cvu}
For any continuous $H:(U_r,d)\to\R$ and any $S \in \phi({\mathbb T}^n)$,
$$
\frac{1}{|B^E_R|} \int_{B^E_R} H(S-t) dt \rightarrow \int_{{\mathbb T}^n} H(\phi(u))du, 
\quad \mbox{as}\ R\rightarrow \infty.
$$
Moreover, the above convergence is uniform with respect to $S \in \phi({\mathbb T}^n)$.
\end{prop}

\begin{prop} \label{dms-distance}
Let $S = \phi(u)$ and $S'=\phi(u')$. Then
$$\od^c(S,S')=\int_{{\mathbb T}^n} d(\phi(u+t),\phi(u'+t))dt.$$
\end{prop}

\begin{prop} \label{dms-continue} The map $\phi$ is continuous from ${\mathbb T}^n$ to $(U_r, \od)$.
\end{prop}
From Proposition \ref{dms-cvu}  and Lemma \ref{ac-plouf}, all sets $S \in \phi({\mathbb T}^n)$ admit a
unique autocorrelation.
From Proposition \ref{dms-continue}, $\phi({\mathbb T}^n)$ is totally bounded in $(U_r,\od)$. 
From Theorem \ref{pp-relcomp}, every $S$ in $\phi({\mathbb T}^n)$ is almost periodic. 
Using Theorem \ref{caract-qc}, one finally gets:

\begin{theoreme} All sets in $\phi({\mathbb T}^n)$ are Patterson sets.
\end{theoreme}

\subsection{Probabilistic results}

We do not assume any more that $|\partial W|_F=0$.
The function $\phi$ defines a stationary and uniformly discrete point process.
For every Borel subset $A\subset {\mathbb T}^n$ and for every $\epsilon>0$, the set
$$\left\{t \in E : |A\cap (A-t)|\geq |A|-\epsilon\right\}$$
is relatively dense, where $|\cdot|$ is the Haar measure on ${\mathbb T}^n$. Hence, $\phi$ fulfills
Assertion~5 of Theorem \ref{ipalmat}. 
Using Theorem \ref{ipalmpatterson}, one finally gets:

\begin{theoreme} For almost all $u \in {\mathbb T}^n$, $\phi(u)$ is a Patterson set.
\end{theoreme}

\bibliographystyle{plain}

\end{document}